\documentstyle[pra,aps,epsfig,graphicx]{revtex}

\def\R{\rm l\!R\,}

\def\ep {\epsilon}

\begin{document}


\title{On The Influence of Two Parallel Plates on Atomic Levels}

\author{D. T. Alves$^{(1)}$\cite{email1}, F. Barone$^{(2)}$\cite{email2},
C. Farina$^{(2)}$\cite{email3} and A. C.
Tort$^{(2)}$\cite{email4}}

\address{${(1)}$ - Universidade Federal do Par\'a, Departamento de F\'{\i}sica,
Bel\'em, PA, CP 479, 66075-110, Brazil}

\address{${(2)}$ - Instituto de F\'{\i}sica,
UFRJ, Caixa Postal 68528,
21945-970 Rio de Janeiro RJ, Brazil}

\date{\today}
\maketitle

\begin{abstract}
This paper is devoted to the study of the influence of two
parallel  plates on the atomic levels of a Hydrogen atom placed in
the region between the plates. We treat two situations, namely:
the case where both plates are infinitely permeable and the case
where one of them is a perfectly conducting plate and the other,
an infinitely permeable one. We compare our result with those
found in literature for two parallel conducting plates. The
limiting cases where the atom is near a conducting plate and near
a permeable one are also taken.
\end{abstract}

\pacs{PACS numbers: 12.20.-m; 32.80.-t} \vskip2pc
\label{sec:level1}


\section{\bf Introduction}

    It has been known for a long time that the consideration of boundary conditions in the radiation field imposed, for instance, by the presence of material plates, may alter not only the vacuum energy, as it occurs in the Casimir effect\cite{Casimir48}, but also the properties of atomic systems in interaction with this field. The most common examples are the influence of cavities on the spontaneous emission rate, or on atomic energy levels (Lamb shift modification) or even on the anomalous magnetic moment of the electron ($g-2$ factor). In other words, we can say that the presence of material walls in the vicinity of atomic systems renormalizes their transition frequencies as well as the widths of their spectral lines. The branch of physics which is concerned with the influence of the environment of atomic systems in their radiative properties is usually called Cavity QED  and the above examples represent only a few of them (for a review see for instance ref(s) \cite{Kleppner89,Haroche,Berman,Milonnilivro}).

    In this paper we shall investigate how the energy levels of a Hydrogen-like atom are altered when it is placed in a region between two parallel plates, where at least one of them is an infinitely permeable plate. These energy modifications are originated from the interaction between the atom and the electromagnetic vacuum fluctuations distorted by the presence of the plates, as pointed out by Power \cite {Power66} in 1966. For the case of perfectly conducting plates, this problem was firstly discussed by Barton a long time ago \cite {Barton70} and later on by L\"utken and Ravndal \cite {LutkenRavndal85}. For the particular case where only one plate is present, the interested reader can consult ref(s) \cite{LutkenRavndal83,PowerThirunamachandran82}. More recently some generalizations were made by Barton \cite {Barton87a,Barton87b}, and Jhe and Nha \cite{jhe90,jhe91}. Cavity QED between parallel dielectric surfaces has also been discussed in the literature \cite{jhe96}.

However, although the influence of permeable plates in the
spontaneous emission rate has already been considered in the
literature \cite{DaniloFarinaTort2000}, its influence in the
atomic energy levels has not, at least as far as the authors'
knowledge. Our purpose here is to fill this gap in the literature.
For simplicity, we shall consider the following situations: {\bf
(i)} a perfectly conducting plate and an infinitely permeable one, wich we will refer to as ({\bf CP}) configuration, 
and {\bf (ii)} two infinitely permeable parallel plates, wich we will refer to as ({\bf PP}) configuration. The
former set-up, used for the first time by Boyer in order to
compute the Casimir effect in the context of stochastic
electrodynamics \cite {Boyer74}, is particularly interesting since
it leads to  a repulsive Casimir pressure \cite {Boyer74} (see
also \cite {Hushwater,Cougo-PintoBJP} and Tenorio {\it et al}
\cite{TenorioTortSantos} for the thermal corrections to this
problem). More recently, the influence of this unusual pair of
plates was also considered in the context of the Scharnhost
effect\cite{Cougo-Pinto98,Cougo-Pinto99}. Regarding the latter
set-up, although it leads to the same Casimir effect as the usual
case (two conducting plates), its influence on the radiative
properties of atomic systems are different.

       In order to calculate the desired energy shifts we shall use second order perturbation theory, regularizing the relevant field correlations with the aid of Schwinger's imaginary time splitting, as in ref.\cite{LutkenRavndal85}. The results are compared with the usual case where both plates are perfect conductors, set-up that, from now on, we shall refer to as ({\bf CC}) configuration.

From these results we shall obtain the energy shifts for an atom
placed near one single conductor plate, and near an infinitlly
permeable one.

\section {\bf Plates with different nature (CP configuration)}
\label {Boyer}

    Let us start by considering the case where the atom is placed between a perfectly conducting plate, located at $z=0$, and an infinitely permeable one, located at $z=L$ (as said, from now on we shall refer to this set-up as ({\bf CP}) configuration). For this case, the corresponding boundary conditions are:
\begin {equation}
\cases{\hat{\bf z}\times{\bf E}(x,y,0,t)=0\cr \hat{\bf z}\cdot{\bf
B}(x,y,0,t)=0}\;\;\;\;\;\;\;\;\;\;\;\;\; \cases{\hat{\bf
z}\times{\bf B}(x,y,L,t)=0\cr \hat{\bf z}\cdot{\bf
E}(x,y,L,t)=0}\;\; .
\end {equation}
In the Coulomb gauge ($\bf\nabla.A=0$) with $A^{0}=0$, we have:
\begin {equation}
{\bf E}=-\dot{\bf A}\ \ \ \ ,\ \ \ \ {\bf B}={\bf\nabla}\times{\bf
A}\; .
\end {equation}

    It is convenient to write separately expressions for the  vector potential for the transverse electric (TE) and magnetic (TM) modes in the following way:
\begin {eqnarray}
\label {expansao1} {\bf A}^{TE}_{\bf k}({\bf
x})&=&{\bf\nabla}\times{\bf U}^{TE}_{\bf k}({\bf x})\cr\cr {\bf
A}^{TM}_{\bf k}({\bf x})&=&{\bf\nabla}\times[{\bf\nabla}\times{\bf
U}^{TM}_{\bf k}({\bf x})],
\end {eqnarray}
where we defined:
\begin {eqnarray}
{\bf U}^{TE}_{\bf k}({\bf x})&=&N{\bf e}_{z}\sin(kz)e^{i{\bf
k}_{T}.{\bf x}}\cr\cr {\bf U}^{TM}_{\bf k}({\bf x})&=&{N\over
i\omega_{\bf k}}{\bf e}_{z}\cos(kz)e^{i{\bf k}_{T}.{\bf x}}
\end {eqnarray}
and used the following notation: ${\bf x}={\bf x}_{T}+z\,\hat{\bf
z}$ .
The wave vector is given by
\begin {equation}
{\bf k}=({\bf k}_{T},k_{z})=(k_{x},k_{y},k_{z})\ ,\
k_{z}={(n+1/2)\pi\over L}=:k\ ,\;\;\;\;\ \cases{k_x,k_y\in \R\cr
n=0,1,2..}
\end {equation}
and hence, the corresponding frequencies read:
\begin {equation}
\omega_{\bf k}=\biggl[{\bf
k}_{T}^{2}+\biggl((n+1/2)\pi/L\biggl)^{2}\biggl]^{1/2} \;\;  .
\end{equation}

    The normalization constant $N=\sqrt{2/{\bf k}_{T}^{2}L}$ is obtained form the condition:
\begin {equation}
\int d^{3}x {\bf A}_{\bf k}^{\lambda *}({\bf x}).{\bf A}_{{\bf
k}'}^{\lambda'}({\bf
x})=4\pi^{2}\delta_{\lambda\lambda'}\delta_{nn'}\delta({\bf
k}_{T}-{\bf k}_{T}')\; .
\end {equation}
Therefore, we can write the vector potential between the plates
as:
\begin {equation}
\label {Aexpandido} {\bf A}({\bf
x})=\sum_{\lambda=E,M}\sum_{n=0}^{\infty}\int{{d^{2}k_{T}\over
(2\pi)^{2}}{1\over\sqrt{2\omega_{\bf k}}}\biggl[a_{\bf
k}^{\lambda}{\bf A}_{\bf k}^{\lambda}({\bf x})e^{-i\omega_{\bf
k}t}+h.c\biggr]}\; ,
\end {equation}
where the anihilation and creation operators satisfy the well
known commutation relations:
\begin {equation}
\biggl[a_{{\bf k}'}^{\lambda'},a_{\bf
k}^{{\lambda}^{\dagger}}\biggr]=4\pi^{2}\delta_{\lambda\lambda'}\delta_{nn'}\delta({\bf
k}_{T}-{\bf k}'_{T}) \; ,
\end {equation}
with all other commutators being zero.

    Our purpose here is to study the effect of the vacuum field fluctuations on the energy levels of an atom placed in a region between the plates. With this goal, we shall use  perturbation theory and assume that the fields do not vary appreciably in the atomic scales (dipole approximation). The first non-vanishing contributions to the energy shifts are obtained in second order in $e$ from:
\begin {equation}
\label {shift1} \Delta\varepsilon_{n}=e^{2}\sum_{m;{\bf
k},\lambda}{{\bf |} \langle n,0|{\bf r}\cdot{\bf E}({\bf
x})|m;{\bf k},\lambda\rangle{\bf
|}^{2}\over\varepsilon_{n}-\varepsilon_{m}-\omega_{\bf k}}\; ,
\end {equation}
with $|n;{\bf k},\lambda\rangle:=\vert n\rangle\otimes\vert{\bf
k},\lambda\rangle$, where $\vert n\rangle$ designates an atomic
state with energy $\varepsilon_{n}$ and $|{\bf k},\lambda\rangle$,
a field state with one photon with momentum $\bf k$ and
polarization $\lambda$. It can be shown that the perturbation
caused by the magnetic field can be
neglected\cite{LutkenRavndal85}. In the above expression, ${\bf
r}$ is the electron position operator of the atom with the origin
taken in its nucleus.

    Separating the contributions for the energy shift (\ref {shift1}) due to the degenerate states and non-degenerate states, denoted respectively by $|n'\rangle$ and $|\ell\rangle$ we write:
\begin {equation}
\label {Deltavarepsilon1}
\Delta\varepsilon_{n}^{(1)}=-e^{2}\sum_{n';{\bf
k},\lambda}{{\bf |}\langle n,0|{\bf r}\cdot{\bf E}({\bf x})|n',{\bf
k},\lambda\rangle{\bf |}^{2}\over\omega_{\bf k}}
\end {equation}
\begin {equation}
\label {Deltavarepsilon2}
\Delta\varepsilon_{n}^{(2)}=e^{2}\sum_{\ell;{\bf k},\lambda}{{\bf
|}\langle n,0|{\bf r}\cdot{\bf E}({\bf x})|\ell;{\bf
k},\lambda\rangle{\bf
|}^{2}\over\varepsilon_{n}-\varepsilon_{\ell}-\omega_{\bf k}}\; .
\end {equation}

    Using selection rules valid for central potentials, or properties of the vacuum field, the energy contribution coming from the degenerated atomic states $\Delta\varepsilon_{n}^{(1)}$ can be written as:
\begin {equation}
\Delta\varepsilon_{n}^{(1)}=-e^{2}\sum_{i}\sum_{n'}{\bf |}\langle
n|x_{i}|n'\rangle{\bf |}^{2}\sum_{{\bf k},\lambda}{{\bf |}\langle
0|E_{i}|{\bf k},\lambda\rangle{\bf |}^{2}\over\omega_{\bf k}}\; .
\end {equation}

    Employing Schwinger's method of imaginary  time splitting we obtain (see the Appendix)
\begin {eqnarray}
\label {corrExEyEz1}
\sum_{{\bf k},\lambda}{{\bf |}\langle
0|E_{x}|{\bf k},\lambda\rangle{\bf |}^{2}\over\omega_{\bf
k}}=\sum_{{\bf k},\lambda}{{\bf |}\langle 0|E_{y}|{\bf
k},\lambda\rangle{\bf |}^{2}\over\omega_{\bf k}}&=&{1\over
2}\sum_{{\bf k},\lambda}{{\bf |}\langle 0|{\bf E}_{T}|{\bf
k},\lambda\rangle{\bf |}^{2}\over\omega_{\bf k}}\cr\cr
&=&{1\over 512L^{3} \pi}G_{+}(z)\cr\cr\cr \sum_{{\bf k},\lambda}{{\bf
|}\langle 0|E_{z}|{\bf k},\lambda\rangle{\bf
|}^{2}\over\omega_{\bf k}}&=&{1\over 256L^{3} \pi}G_{-}(z).
\end {eqnarray}
where we defined:
\begin {eqnarray}
G_{\pm}(z)&=&\zeta_{H}(3,z/2L)+\zeta_{H}(3,-z/2L)-\zeta_{H}(3,1/2+z/2L)+\cr\cr
&-&\zeta_{H}(3,1/2-z/2L)+\biggl({2L\over z}\biggr)^{3}\pm
12\zeta_{R}(3)
\end {eqnarray}
with $\zeta_{H}$ and $\zeta_{R}$ being the Huruwitz and Reimann
zeta functions, respectively. As a consequence, the contribution
coming from the degenerated levels to the energy shifts are given
by:
\begin {eqnarray}
\label {Deltavarepsilon(1)dist1}
\Delta\varepsilon_{n}^{(1)}=-{e^{2}\over 512\pi
L^{3}}\sum_{n'}\biggl[({\bf |}\langle n|x|n'\rangle{\bf
|}^{2}+{\bf |}\langle n|y|n'\rangle{\bf |}^{2})G_{+}(z)+2{\bf
|}\langle n|z|n'\rangle{\bf |}^{2}G_{-}(z)\biggr]\; .
\end {eqnarray}

    Let us now address our attention to the contribution coming from the non-degenerate levels. For this case, we shall consider separately two limiting regimes, namely, where the atom is near one plate and when it is far away from the plates.

    Near one of the plates, it can be shown that the dominant contribution  comes from $\omega_{\bf k}>>\varepsilon_{n}-\varepsilon_{\ell}$ \cite{LutkenRavndal85}. Hence, neglecting $\varepsilon_{n}-\varepsilon_{\ell}$ and using, as before, arguments based on selection rules for atomic transitions with spherical symmetric potentials, or properties of the vacuum fields, it can be shown that the two contributions  $\Delta\varepsilon_{n}^{(1)}$ and $\Delta\varepsilon_{n}^{(2)}$ (Eq's (\ref {Deltavarepsilon1}) and (\ref {Deltavarepsilon2})) take the same form. Consequently, using the completeness of the atomic states we get:
\begin {eqnarray}
\label {finalBoyerperto}
\Delta\varepsilon_{n}&=&\Delta\varepsilon_{n}^{(1)}+\Delta\varepsilon_{n}^{(2)}\cr\cr
&=&-{e^{2}\over 512\pi L^{3}}\biggl[\langle
n|(x^{2}+y^{2})|n\rangle G_{+}(z)+2\langle n|z^{2}|n\rangle
G_{-}(z)\biggr]
\end {eqnarray}

\

    Far away from the plates (retarded regime), it can be shown that the dominant contribution comes from $\omega_{\bf k}<<\varepsilon_{n}-\varepsilon_{\ell}$ \cite{LutkenRavndal85}. Discarding now $\omega_{\bf k}$, the contribution $\Delta\varepsilon_{n}^{(2)}$ becomes:
\begin {equation}
\Delta\varepsilon_{n}^{(2)}=e^{2}\sum_{i}\sum_{\ell}{{\bf
|}\langle n|x_{i}|\ell\rangle{\bf
|}^{2}\over\varepsilon_{n}-\varepsilon_{\ell}}\sum_{{\bf
k},\lambda}{\bf |}\langle 0|{\bf E}_{i}|{\bf k},\lambda\rangle{\bf
|}^{2}.
\end {equation}
Using the definition of the static electric polarizabilities of
level n,
\begin {equation}
\alpha_{i}\equiv 2e^{2}\sum_{\ell}{{\bf |}\langle
n|x_{i}|\ell\rangle{\bf
|}^{2}\over\varepsilon_{\ell}-\varepsilon_{n}}
\end {equation}
as well as the matrix elements of the electric field operator
obtained in the Appendix, we have in the diagonal basis of atomic
states:
\begin {equation}
\label {finalBoyerlongeND} \label {Deltavarepsilon(2)dist1}
\Delta\varepsilon_{n}^{(2)}=-{\pi^{2}\over
96L^{4}}\biggl[(\alpha_{x}+\alpha_{y}+\alpha_{z})\biggl({G(\theta)\over
2}+{7\over 720}\biggr)-\alpha_{z}{7\over 360}\biggr]\; ,
\end {equation}
where we defined:
\begin{equation}
G(\theta)=6{\cos\theta\over\sin^{4}\theta}-{\cos\theta\over\sin^{2}\theta}\;
.
\end{equation}
 To have the total shift away from the plates we must consider the contributions $\Delta\varepsilon_{n}^{(1)}$ and $\Delta\varepsilon_{n}^{(2)}$ of equations (\ref {Deltavarepsilon(1)dist1}) and (\ref {Deltavarepsilon(2)dist1}).

\section {\bf Two infinitely permeable plates (PP configuration)}

    Comsidering now the ({\bf PP}) configuration, that is, two infinitely permeable parallel plates, the boundary conditions on the electromagnetic fields are now given by:
\begin {equation}
\cases{\hat{\bf z}\times{\bf B}(x,y,0,t)=0\cr \hat{\bf z}\cdot{\bf
E}(x,y,0,t)=0}\;\;\;\;\;\;\;\;\;\;\;\;\; \cases{\hat{\bf
z}\times{\bf B}(x,y,L,t)=0\cr \hat{\bf z}\cdot{\bf
E}(x,y,L,t)=0}\;\; .
\end {equation}
Using the same gauge as before (${\bf\nabla\cdot A}=0$ with
$A^{0}=0$), and writing separately the vector potential for the TE
and TM modes, as we did for the ({\bf CP}) configuration (see section
\ref {Boyer}), we have:
\begin {eqnarray}
{\bf U}^{TE}_{\bf k}({\bf x})&=&N{\bf e}_{z}\cos(kz)e^{i{\bf
k}_{T}.{\bf x}}\cr\cr {\bf U}^{TM}_{\bf k}({\bf x})&=&{N\over
i\omega_{\bf k}}{\bf e}_{z}\sin(kz)e^{i{\bf k}_{T}.{\bf x}}
\end {eqnarray}
\begin {equation}
k={n\pi\over L}\ ,  \ n=0,1,2...\ \ \ \ \ k_x,k_y\in\R\ \ \ \   ,
\end {equation}

    Following the same procedure as that employed for the ({\bf CP}) configuration, we obtain after a lengthy but straightforward calculation, that the energy shifts when the atom is near one of the plates are obtained from:
\begin {equation}
\label {finalPPperto} \Delta\varepsilon_{n}={e^{2}\over 64\pi
L^{3}}\biggl[\langle n|(x^{2}+y^{2})|n\rangle F_{+}(z)+2\langle
n|z^{2}|n\rangle F_{-}(z)\biggr]\; ,
\end {equation}
where we defined:
\begin {equation}
\label{defFmaismenos}
F_{\pm}(z)=\zeta_{H}(3,z/L)+\zeta_{H}(3,-z/L)\pm
2\zeta_{R}(3)+\biggl({L\over z}\biggr)^{3}\; .
\end {equation}

    Away from the plates, the contributions coming from the degenarated states to the energy shifts are obtained from:
\begin {equation}
\label {finalPPlongeDEG} \Delta\varepsilon_{n}^{(1)}={e^{2}\over
64\pi L^{3}}\sum_{n'}\Biggl[\biggl({\bf |}\langle
n|x|n'\rangle{\bf |}^{2}+{\bf |}\langle n|y|n'\rangle{\bf
|}^{2}\biggr)F_{+}(z)+2{\bf |}\langle n|z|n'\rangle{\bf
|}^{2}F_{-}(z)\Biggr]
\end {equation}
and the contributions from the non degenerated states, in a basis
that diagonalizes the atomic states, read:
\begin {equation}
\label {finalPPlongeNDEG}
\Delta\varepsilon_{n}^{(2)}={\pi^{2}\over
96L^{4}}\Biggl[(\alpha_{x}+\alpha_{y}+\alpha_{z})\Biggl(F(\theta)+{1\over
15}\Biggr)-{2\over 15}\alpha_{z}\Biggr]
\end {equation}
where:
\begin {equation}
\label{defF(theta)}
F(\theta)={3\over\sin^{4}(\theta)}-{2\over\sin^{2}(\theta)}\ .
\end {equation}

\section {\bf Comments and Conclusions}

    This section is devoted to compare the results obtained in this paper, and the results presented in reference \cite {LutkenRavndal85} where we have the ({\bf CC}) configuration, that is, two conductors plates. For this boundary condition, the energy shifts of an atom placed near one of the plates come from:
\begin {equation}
\label {finalCasimirperto} \Delta\varepsilon_{n}=-{e^{2}\over
64\pi L^{3}}\biggl[\langle n|(x^{2}+y^{2})|n\rangle
F_{-}(z)+2\langle n|z^{2}|n\rangle F_{+}(z)\biggr].
\end {equation}

    The contributions to the shifts due to the degenerated states come from:
\begin {equation}
\label {finalCasimirlongeDEG}
\Delta\varepsilon_{n}^{(1)}=-{e^{2}\over 64\pi
L^{3}}\sum_{n'}\Biggl[\biggl({\bf |}\langle n|x|n'\rangle{\bf
|}^{2}+{\bf |}\langle n|y|n'\rangle{\bf |}^{2}\biggr)F_{-}(z)
+2{\bf |}\langle n|z|n'\rangle{\bf |}^{2}F_{+}(z)\Biggr]
\end {equation}
for any distance from the plates.

    Far away from the plates, the enegy shifts contributions coming from the non degenerated states, in a basis that diagonalizes the atomic states, read:
\begin {equation}
\label {finalCasimirlongeNDEG}
\Delta\varepsilon_{n}^{(2)}=-{\pi^{2}\over
96L^{4}}\Biggl[(\alpha_{x}+\alpha_{y}+\alpha_{z})\Biggl(F(\theta)-{1\over
15}\Biggr)+{2\over 15}\alpha_{z}\Biggr]
\end {equation}
with the functions $F(\theta)$ and $F_{\pm}(z)$ defined in ({\ref
{defF(theta)}) and ({\ref{defFmaismenos}) respectively.

It is interesting to note that the function $F(\theta)$ is
strictlly positive along its domain (see reference
\cite{LutkenRavndal85}), but the function $G(\theta)$ can be
positive or negative, as shown in figure \ref{figure1}, which
gives completely different behaviours for the energy shifts
contributions (\ref {finalPPlongeNDEG}) and (\ref
{finalCasimirlongeNDEG}) in comparing with (\ref
{finalBoyerlongeND}).

In order to do a numerical analysis of our results, let's restrict
ourselves, from now on, to atoms not too highly excited. In this
case it can be shown \cite {LutkenRavndal85} that the
contributions coming from the non degenarated atomic states,
(\ref{finalCasimirlongeDEG}), (\ref{finalPPlongeDEG}) and
(\ref{Deltavarepsilon(1)dist1}), are the relevant to the energy
shifts. Their signs are determined from the signs of the functions
$F_{\pm}(z)$ and $G_{\pm}(z)$; the formers are strictly positive
and the latters can change their signs, giving  negative shifts
for the ({\bf CC}) configuration, while positive shifts for ({\bf PP}) configuration. Note from equations (\ref {finalCasimirlongeDEG}) and  (\ref{finalPPlongeDEG}) that these
contributions to the energy shifts have opposite signs. Further,
the roles of the longitudinal and transverse field fluctuations
are also interchanged. For ({\bf CP}) configuration, the shifts can be
positive or negative.

Here we can point out some differences between these energy shifts
and the Casimir effect, another important manifestation of the
vacuum fluctuations. For the Casimir effect, the ({\bf PP})
and ({\bf CC}) plates give the same attractive Casimir
force, while for ({\bf CP}) plates we have a repulsive Casimir force
(but with the same $L$ dependence as for the other two boundary
conditions). In contrast,  for the energy shifts we expect
different behaviours even for those cases where the Casimir
energies are the same, since the atom probes locally the quantum
vacuum fluctuations, while the Casimir energy is a global
quantity.

Now we present a table with numerical results, showing the energy
shifts computed for the lowest hydrogen levels when the atom
interacts with the radiation field in vacuum state submitted to
the three boundary conditions mentioned above. For simplicity, we
assume that the atom is placed at the midle point between the
plates ($z=L/2$). The results are in units of $\zeta_{R}(3)(\alpha
m)^{-2}\alpha/32L^{3}$ (the results for ({\bf CC}) plates can be
found in reference \cite {LutkenRavndal85}).

\begin {equation}
\begin{tabular}{|c|c|c|c|}
\hline \mbox{Energy Shifts} & \mbox{Casimir} & \mbox{double
permeable} & \mbox {Boyer}  \\
\hline $\Delta\ep^{(1)}_{200}$ & -1008                &  1008
& 0   \\
\hline $\Delta\ep^{(1)}_{210}$ & -576                 &  432
& -54  \\
\hline $\Delta\ep^{(1)}_{211}$ & -216                 &  288
& 27   \\
\hline $\Delta\ep^{(1)}_{+}$   & -162(25+$\sqrt 241$) &
1296(3+$\sqrt 3$) & -60,75(1+$\sqrt 33$)   \\
\hline $\Delta\ep^{(1)}_{-}$   & -162(25-$\sqrt 241$) &
1296(3-$\sqrt 3$) & -60,75(1-$\sqrt 33$)   \\
\hline $\Delta\ep^{(1)}_{310}$ & -6156                &  5184
&  -364,5   \\
\hline $\Delta\ep^{(1)}_{311}$ & -3726                &  4212
&  182,2   \\
\hline $\Delta\ep^{(1)}_{321}$ & -1782                &  1620
&  -60,75   \\
\hline $\Delta\ep^{(1)}_{322}$ & -972                 &  1296
&  121,5   \\
\hline
\end{tabular}
\end {equation}

We can see from the above table that the energy shifts  for the
({\bf CC}) and ({\bf PP}) plates are of the same order, while
for ({\bf CP}) plates they are one order of magnitude smaller. Remark
that for the Casimir effect, ({\bf CP}) configuration also leads to
a smaller force, in modulus, than ({\bf CC}) original
configuration, but they are of the same order of magnitude.

As a last comment, let us analyse the limiting cases where the
atom is near a unique perfectly conducting plate as well as a
unique perfectly permeable one. For the former case we take in
equation (\ref{finalBoyerperto}) the limit of the atom located
near the conducting plate, namely, we just make $z/L\rightarrow0$:
\begin {equation}
\label {1placaC} \Delta\varepsilon_{(n)}=-{e^{2}\over 64\pi
z^{3}}\biggl[\langle n|(x^{2}+y^{2})|n\rangle +2\langle
n|z^{2}|n\rangle \biggr]
\end {equation}

This same result can be obtained from the Lutken and Ravndal's
paper \cite{LutkenRavndal85}.

For the second case, it would be better for calculations to have a
formula that gives the energy shifts for an infinitely permeable
plate at $z=0$, and a perfectly conductor one at $z=L$. This
expression can be obtained making the substitution
$z\rightarrow-(z-L)$ in equation (\ref{finalBoyerperto}):
\begin {equation}
\Delta\varepsilon_{n}=-{e^{2}\over 512\pi L^{3}}\biggl[\langle
n|(x^{2}+y^{2})|n\rangle G_{+}(-(z-L))+2\langle n|z^{2}|n\rangle
G_{-}(-(z-L))\biggr]\; .
\end {equation}
We can then take the limit $z/L\rightarrow0$, giving for an atom
near one infinitely permeable plate the energy shifts:
\begin {equation}
\label {1placaP} \Delta\varepsilon_{(n)}={e^{2}\over 64\pi
z^{3}}\biggl[\langle n|(x^{2}+y^{2})|n\rangle +2\langle
n|z^{2}|n\rangle \biggr]
\end {equation}

We could also have obtained this  expression taking the limit
$z/L\rightarrow0$ in equation (\ref{finalPPperto}) as well. Note
that the energy shifts for one conducting plate (\ref {1placaC})
and one perfectly permeable one (\ref {1placaP}) have opposite
signs and same magnitude.

As a last comment, we would like to emphasize the increasing
importance of considering the influence of permeable plates in
different physical situations \cite{KlichPRD2002,KennethPRL2002}
(see also references therein). This is clear, for instance, if we
note that Casimir forces may become dominant at the nonometer
scale and the appropriate consideration of permeable plates can
produce repulsive forces. Recall that only attractive forces could
lead to restrictive limits on the construction of nanodevices.

\appendix
\section*{}

Using the plane wave expansion of the vector potential (\ref
{Aexpandido}), we can easily show that:

\begin {equation}
\label {soma->valvacuo} \sum_{{\bf k},\lambda}{{\bf |}\langle
0|E_{i}|{\bf k},\lambda\rangle{\bf |}^{2}}=\langle
0|E_{i}E_{i}|0\rangle\ .
\end {equation}
These correlators are plagued with infinities and must be
regularized. Choosing the Schwinger's imaginary time splitting, we
write the regularized transverse and longitudinal vacuum
fluctuations of the electric field operator respectivelly as:
\begin {eqnarray}
\label {flutuacoes1} \langle 0|{\bf
E}_{T}^{2}(z)|0\rangle&=&\lim_{t'\rightarrow t}\langle 0|{\bf
E}_{T}(z,t){\bf E}_{T}(z,t')|0\rangle\cr\cr &=&{1\over
L}\lim_{t'\rightarrow t}\sum_{n=0}^{\infty}\int{d^{2}{\bf
k}_{T}\over(2\pi)^{2}}\biggl(\omega_{\bf k}+{k^{2}\over\omega_{\bf
k}}\biggr)\sin^{2}(kz)e^{i\omega(t'-t)}\cr\cr \langle
0|E_{z}^{2}(z)|0\rangle&=&\langle
0|E_{z}(z,t)E_{z}(z,t')|0\rangle\cr\cr &=&{1\over
L}\lim_{t'\rightarrow t}\sum_{n=0}^{\infty}\int{d^{2}{\bf
k}_{T}\over(2\pi)^{2}}\biggl(\omega_{\bf k}-{k^{2}\over\omega_{\bf
k}}\biggr)\cos^{2}(kz)e^{i\omega(t'-t)}\ ,
\end {eqnarray}
where $t'-t$ will be substitute by $i\tau$ (this is equivalent to
introduce an exponential cut off). Further, defining:
\begin {equation}
\label {deftau...} \lambda={\pi\over L}\ \ \ ,\ \ \
\epsilon=\lambda\tau\ \ \ ,\ \ \ \theta=\lambda z
\end {equation}
and:
\begin {eqnarray}
\label {defXi}
\Xi_{\underline{+}}(\epsilon,\theta)&=&\sum_{n=0}^{\infty}\biggl[1\underline{+}
\cos(2kR)\biggr]\cr\cr &=&\sum_{n=0}^{\infty}\biggl[1\underline{+}
\cos\biggr(2\pi(n+1/2)z/L\biggr)\biggr]\cr\cr &=&{1\over
2\sinh(\epsilon/2)}\underline{+}{1\over
4}\biggl[{1\over\sinh(\epsilon/2 -i\theta)}+c.c\biggr]
\end {eqnarray}
we have, omitting the limit $\tau\rightarrow 0$:
\begin {eqnarray}
\langle 0|{\bf E}_{T}^{2}(z)|0\rangle&=&{\lambda^{3}\over 4\pi
L}\biggl(\partial_{\epsilon}{1\over\epsilon}+{1\over\epsilon}\partial_{\epsilon}\biggr)\Xi_{-}(\epsilon,\theta)\cr\cr
\langle 0|E_{z}^{2}(z)|0\rangle&=&{\lambda^{3}\over 4\pi
L}\biggl(\partial_{\epsilon}{1\over\epsilon}-{1\over\epsilon}\partial_{\epsilon}\biggr)\Xi_{+}(\epsilon,\theta).
\end {eqnarray}
Expanding Eq.(\ref {defXi}) in powers of $\epsilon$, that is:
\begin {equation}
\label {expansao} \Xi_{\pm}={1\over\epsilon}-{\epsilon\over
4}T_{\pm}(\theta)+{\epsilon^{3}\over 96}\biggl({7\over 60}\mp
G(\theta)\biggr)+O(\epsilon^{4})\ ,
\end {equation}
where
\begin {eqnarray}
T_{\pm}(\theta)&=&{1\over
6}\mp{\cos\theta\over\sin^{2}\theta}\cr\cr
G(\theta)&=&6{\cos\theta\over\sin^{4}\theta}-{\cos\theta\over\sin^{2}\theta}\
,
\end {eqnarray}
we obtain:
\begin {eqnarray}
\label {renA} \langle 0|{\bf
E}_{T}^{2}(z)|0\rangle&=&{\pi^{2}\over
4L^{4}}\biggl[{8\over\epsilon^{4}}+{1\over
12}\biggl(G(\theta)+{7\over 360}\biggr)\biggr]\cr\cr \langle
0|E_{z}^{2}(z)|0\rangle&=&{\pi^{2}\over
4L^{4}}\biggl[{4\over\epsilon^{4}}+{1\over
24}\biggl(G(\theta)-{7\over 360}\biggr)\biggr]\ .
\end {eqnarray}

It's not a difficult task to show that
\begin {eqnarray}
\label {ET->Ex,Ey} \langle 0|E_{x}^{2}(z)|0\rangle=\langle
0|E_{y}^{2}(z)|0\rangle&=&{1\over 2}\langle 0|{\bf
E}_{T}^{2}(z)|0\rangle\cr\cr &=&{\pi^{2}\over
4L^{4}}\Biggl[{4\over\epsilon^{4}}+{1\over
24}\Biggl(G(\theta)+{7\over 360}\Biggr)\Biggr]\ .
\end {eqnarray}
The terms proportional to $1/\varepsilon^{4}$ in Eq's (\ref{renA})
and (\ref{ET->Ex,Ey}) diverge in the limit $\tau\rightarrow
0$, however they are $L$-independent, so that they are spurious
terms with no physical significance.

Using the same procedure adopted to compute Eq's (\ref
{soma->valvacuo}) and (\ref {flutuacoes1}), we can write:
\begin {eqnarray}
\sum_{{\bf k},\lambda}{{\bf |}\langle 0|{\bf E}_{T}|{\bf
k},\lambda\rangle{\bf |}^{2}\over\omega_{\bf
k}}&=&\lim_{t'\rightarrow t}\sum_{{\bf k},\lambda}{\langle 0|{\bf
E}_{T}(z,t)|{\bf k},\lambda\rangle\langle{\bf k},\lambda|{\bf
E}_{T}(z,t')|0\rangle\over\omega_{\bf k}}\cr\cr &=&{1\over
L}\lim_{t'\rightarrow t}\sum_{n=0}^{\infty}{\int{d^{2}{\bf
k}_{T}\over(2\pi)^{2}}{1\over\omega_{\bf k}}\biggl[\omega_{\bf
k}+{k^{2}\over\omega_{\bf k}}\biggr]\sin^{2}(kz)e^{i\omega_{\bf
k}(t'-t)}}\cr\cr\cr \sum_{{\bf k},\lambda}{{\bf |}\langle
0|E_{z}|{\bf k},\lambda\rangle{\bf |}^{2}\over\omega_{\bf
k}}&=&\lim_{t'\rightarrow t}\sum_{{\bf k},\lambda}{\langle
0|E_{z}(z,t)|{\bf k},\lambda\rangle\langle{\bf
k},\lambda|E_{z}(z,t')|0\rangle\over\omega_{\bf k}}\cr\cr
&=&{1\over L}\lim_{t'\rightarrow
t}\sum_{n=0}^{\infty}{\int{d^{2}{\bf
k}_{T}\over(2\pi)^{2}}{1\over\omega_{\bf k}}\biggl[\omega_{\bf
k}-{k^{2}\over\omega_{\bf k}}\biggr]\cos^{2}(kz)e^{i\omega_{\bf
k}(t'-t)}}
\end {eqnarray}
With definitions (\ref {deftau...}) and (\ref {defXi}), and
omitting as before the limit $\tau\rightarrow 0$, we
obtain:
\begin {eqnarray}
\label {corr/omega1} \sum_{{\bf k},\lambda}{{\bf |}\langle 0|{\bf
E}_{T}|{\bf k},\lambda\rangle{\bf |}^{2}\over\omega_{\bf
k}}&=&{\pi\over 2L^{3}}\biggl[\int_{\epsilon}^{\infty}dx
g_{-}(x)-{1\over\epsilon}{\partial\over\partial\epsilon}\Xi_{-}(\epsilon,\theta)\biggr]\cr\cr
\sum_{{\bf k},\lambda}{{\bf |}\langle 0|E_{z}|{\bf
k},\lambda\rangle{\bf |}^{2}\over\omega_{\bf k}}&=&{\pi\over
2L^{3}}\biggl[{1\over\epsilon^{2}}\Xi_{+}(\epsilon,\theta)-\int_{\epsilon}^{\infty}dx\,
g_{+}(x)\biggr]
\end {eqnarray}
where $g_{\pm}(x)=(1/x^{3})\Xi_{\pm}(x,\theta)$. In order to
compute the above integrals we consider the analytically continued
function
\begin {equation}
g_{\pm}(z,p)={1\over z^{p}}\Xi_{\pm}(z,\theta)
\end {equation}

The integral along $C_{\rho}$ (see Fig.(2)) vanishes, which
yields, using the residue theorem:
\begin {equation}
\int_{\epsilon}^{\infty}dx g_{\pm}(x,p)=(1-e^{-2\pi
pi})^{-1}\biggl[-\int_{C_{\epsilon}}dz g_{\pm}(z,p)+2\pi
i\sum_{z\not=0}Res g_{\pm}(x,p)\biggr]
\end {equation}

    With the definitions:
\begin {equation}
\label {ultimo} \zeta_R(s)=\sum_{n=1}^{\infty}{1\over n^{s}}\ \ \
\ ,\ \ \ \zeta_H(s,a)=\sum_{n=1}^{\infty}{1\over(n+a)^{s}}
\end {equation}
and using the expansion (\ref {expansao}), we have:
\begin {equation}
\lim_{p\rightarrow 3}\biggl[(1-e^{-2\pi
pi})^{-1}(-1)\int_{C_{\epsilon}}dz
g_{\underline{+}}(z,p)\biggr]=\biggl[{1\over
3\epsilon^{3}}-{T_{\pm}(\theta)\over 4\epsilon}\biggr]
\end {equation}
and

\begin {eqnarray}
&\ \ \ &\lim_{p\rightarrow 3}[(1-e^{-2\pi pi})^{-1}2\pi
i\sum_{z\not=0}Res g_{\underline{+}}(x,p)]=\cr\cr &\
&\;\;\;\;\;\;\;\;\;\;=-{1\over
2(4\pi)^{2}}\Biggl[-6\zeta_{R}(3)\pm{1\over
2}\Biggl(\zeta(3,\theta/2\pi)+\zeta(3,-\theta/2\pi)+\cr\cr &\
&\;\;\;\;\;\;\;\;\;\;-\zeta(3,1/2+\theta/2\pi)-\zeta(3,1/2-\theta/2\pi)+\biggl({2\pi\over\theta}\biggr)^{3}\Biggr)\Biggr]\cr\cr
&\ &\;\;\;\;\;\;\;\;\;\;\equiv \mp{1\over
128\pi^{2}}G_{\mp}(\theta)
\end {eqnarray}
which gives the desired integral:
\begin {eqnarray}
\label {intg pm} \int_{\varepsilon}^{\infty}dx
g_{\pm}(x)&=&\int_{\varepsilon}^{\infty}dx {1\over
x^{3}}\Xi_{\pm}(x,\theta)\cr\cr &=&{1\over
3\varepsilon^{3}}-{T_{\pm}(\theta)\over 4\varepsilon}\mp{1\over
128\pi^{2}}G_{\mp}(\theta)\ .
\end {eqnarray}

Subsituting this result in Eq. (\ref {corr/omega1}), and using the
definition of $\Xi$, we obtain the result (\ref {corrExEyEz1}),
and a term that diverges in the limit $\tau\rightarrow 0$.
As before it is $L$-independent, so that it's a spurious term with
no physical significance.

\clearpage
\begin{figure}[p]
\label{figure1}
\vspace{7.9cm}
\begin{center}
\includegraphics{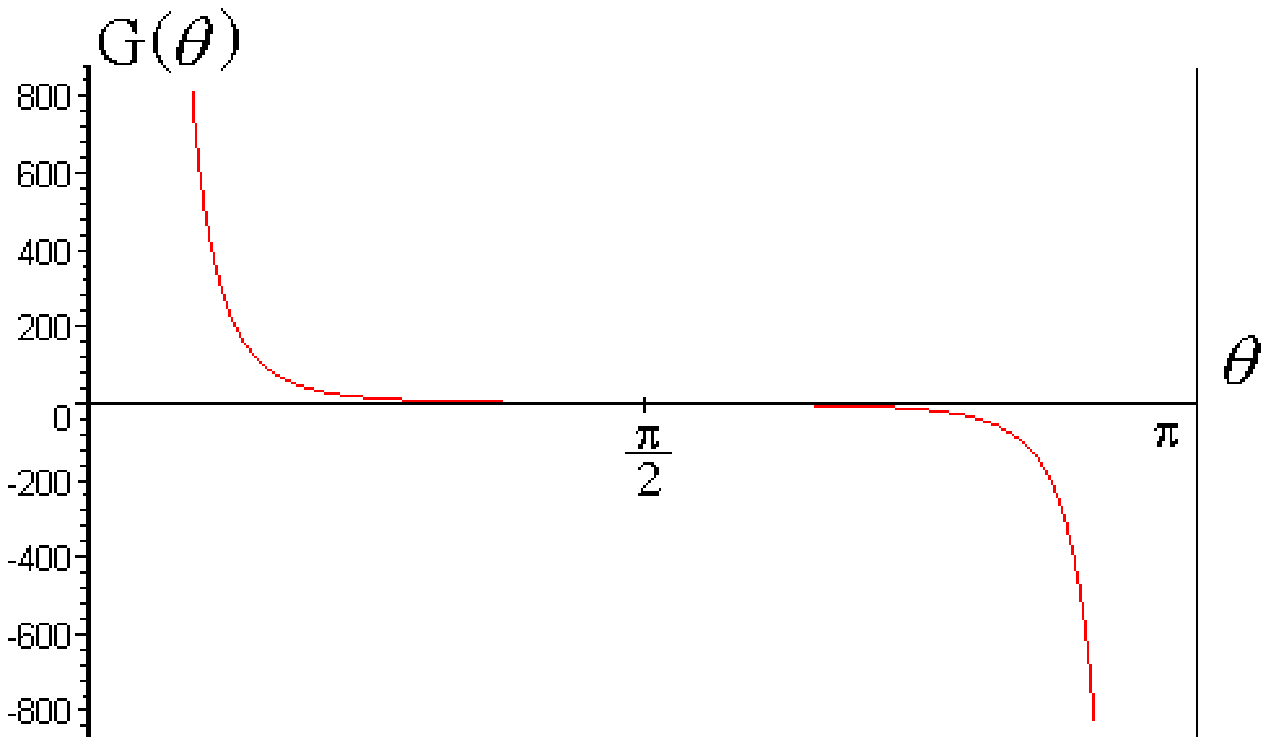}
\end{center}
\caption{Graphic for $G(\theta)$.}
\end{figure}

\begin{figure}[p]
\label{figure2}
\vspace{10.7cm}
\begin{center}
\includegraphics{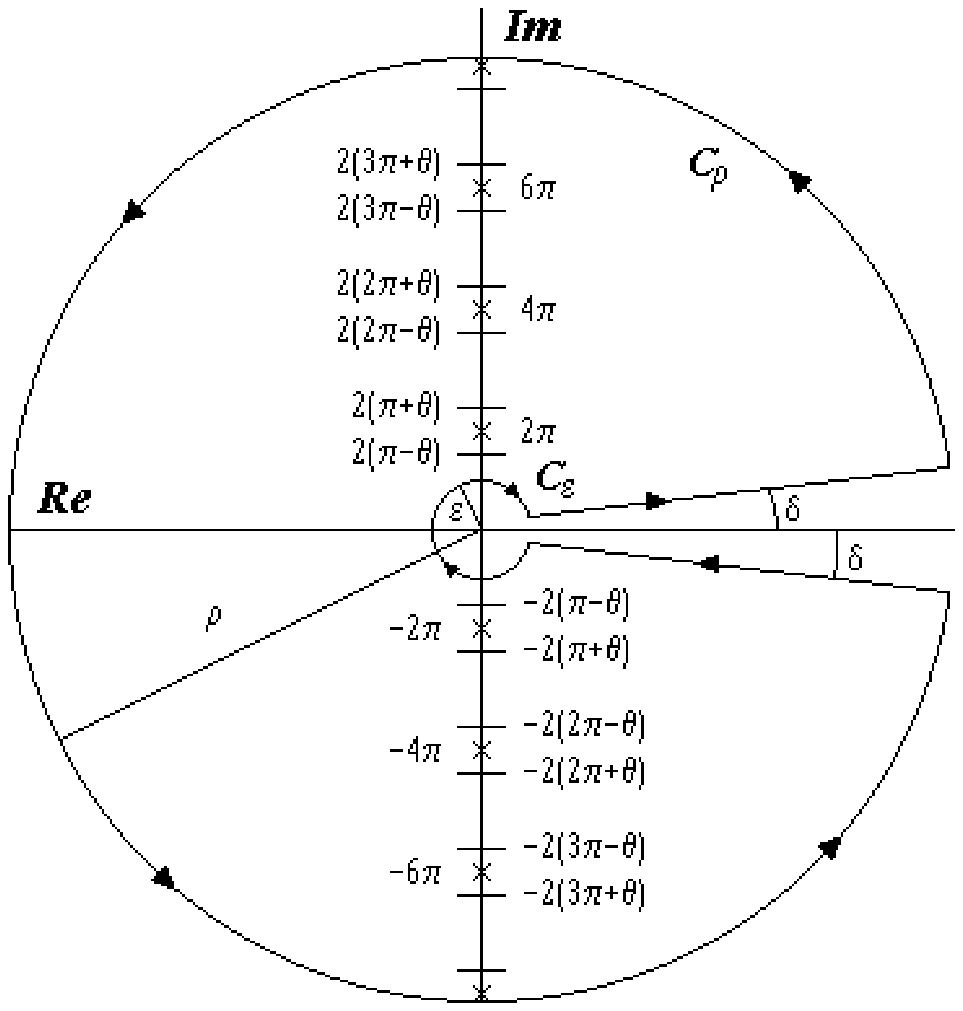}
\end{center}
\caption{Integration contour.}
\end{figure}

\end{document}